\documentclass{emulateapj}

\begin{document}

\title{Investigating the Cosmic-Ray Ionization Rate Near the Supernova Remnant IC 443 Through H$_3^+$ Observations\altaffilmark{1,2}}

\author{Nick Indriolo\altaffilmark{3},
Geoffrey A. Blake\altaffilmark{4},
Miwa Goto\altaffilmark{5},
Tomonori Usuda\altaffilmark{6},
Takeshi Oka\altaffilmark{7},
T. R. Geballe\altaffilmark{8},
Brian D. Fields\altaffilmark{3,9}
Benjamin J. McCall\altaffilmark{3,9,10}}
\altaffiltext{1}{Some of the data presented herein were obtained at the W.M. Keck Observatory, which is operated as a scientific partnership among the California Institute of Technology, the University of California and the National Aeronautics and Space Administration. The Observatory was made possible by the generous financial support of the W.M. Keck Foundation.}
\altaffiltext{2}{Based in part on data collected at Subaru Telescope, which is operated by the National Astronomical Observatory of Japan.}
\altaffiltext{3}{Department of Astronomy, University of Illinois at
Urbana-Champaign, Urbana, IL 61801}
\altaffiltext{4}{Division of Geological and Planetary Sciences and Division of Chemistry and Chemical Engineering, MS 150-21, California Institute of Technology, Pasadena, CA 91125}
\altaffiltext{5}{Max-Planck-Institut f\"{u}r Astronomie, K\"{o}nigstuhl 17, Heidelberg D-69117, Germany}
\altaffiltext{6}{Subaru Telescope, 650 North A'ohoku Place, Hilo, HI 96720}
\altaffiltext{7}{Department of Astronomy and Astrophysics and Department of Chemistry, University of Chicago, Chicago, IL 60637}
\altaffiltext{8}{Gemini Observatory, 670 North A'ohoku Place, Hilo, HI 96720}
\altaffiltext{9}{Department of Physics, University of Illinois at
Urbana-Champaign, Urbana, IL 61801}
\altaffiltext{10}{Department of Chemistry, University of Illinois at
Urbana-Champaign, Urbana, IL 61801}

\begin{abstract}
Observational and theoretical evidence suggests that high-energy Galactic cosmic rays are primarily accelerated by supernova remnants.  If also true for low-energy cosmic rays, the ionization rate near a supernova remnant should be higher than in the general Galactic interstellar medium (ISM).  We have searched for H$_3^+$ absorption features in 6 sight lines which pass through molecular material near IC 443---a well-studied case of a supernova remnant interacting with its surrounding molecular material---for the purpose of inferring the cosmic-ray ionization rate in the region.  In 2 of the sight lines (toward ALS 8828 and HD 254577) we find large H$_3^+$ column densities, $N({\rm H}_3^+)\approx3\times10^{14}$~cm$^{-2}$, and deduce ionization rates of $\zeta_2\approx2\times10^{-15}$~s$^{-1}$, about 5 times larger than inferred toward average diffuse molecular cloud sight lines.  However, the $3\sigma$ upper limits found for the other 4 sight lines are consistent with typical Galactic values.  This wide range of ionization rates is likely the result of particle acceleration and propagation effects, which predict that the cosmic-ray spectrum and thus ionization rate should vary in and around the remnant.  While we cannot determine if the H$_3^+$ absorption arises in post-shock (interior) or pre-shock (exterior) gas, the large inferred ionization rates suggest that IC 443 is in fact accelerating a large population of low-energy cosmic rays.  Still, it is unclear whether this population can propagate far enough into the ISM to account for the ionization rate inferred in diffuse Galactic sight lines.

\end{abstract}


\section{INTRODUCTION}

As cosmic rays propagate through the interstellar medium (ISM) they interact with the ambient material.  These interactions include excitation and ionization of atoms and molecules, spallation of nuclei, excitation of nuclear states, and the production of neutral pions ($\pi^0$) which decay into gamma-rays.  Evidence suggests that Galactic cosmic rays are primarily accelerated by supernova remnants (SNRs) through the process of diffusive shock acceleration \citep[e.g.][]{drury83,blandford87}, so interstellar clouds in close proximity to an SNR should provide a prime ``laboratory'' for studying these interactions.  IC 443 represents such a case, as portions of the SNR shock are known to be interacting with the neighboring molecular clouds.

\begin{figure}
\epsscale{1.0}
\plotone{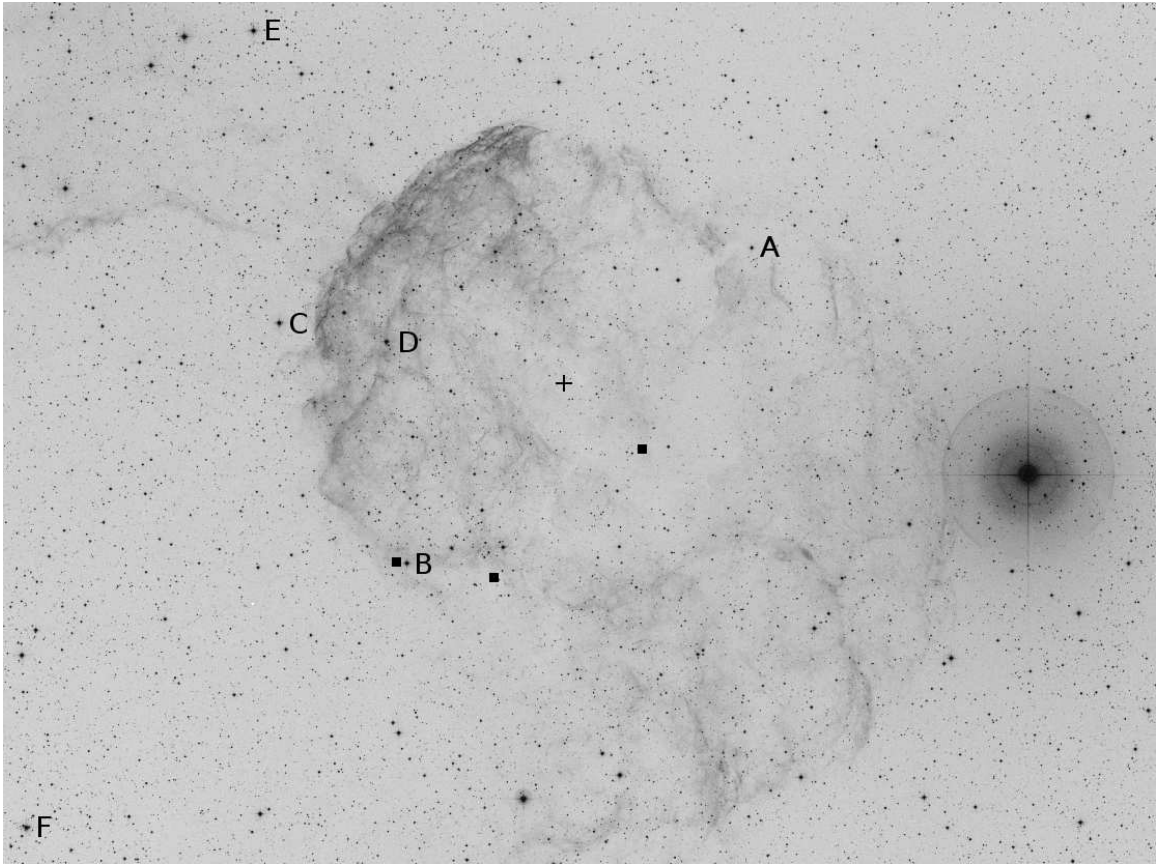}
\caption{This image of IC 443 is from the Second Palomar Observatory Sky Survey (POSS-II) using the red filter, and was obtained from the STScI Digitized Sky Survey.  Target background stars are to the immediate left of the uppercase letters, and are labeled as follows: A-ALS 8828; B-HD 254577; C-HD 254755; D-HD 43582; E-HD 43703; F-HD 43907.  The cross marks the center of subshell A (the shell on the northeast side of IC 443) at $\alpha=06^{\rm h}17^{\rm m}08.4^{\rm s}$, $ \delta=+22^{\circ}36'39.4''$ J2000.0.  The three black squares mark the positions of OH (1720~MHz) maser emission reported by \citet{hewitt06}.}
\label{figIC443}
\end{figure}

IC 443 is an intermediate age remnant \citep[about 30,000 yr;][]{chevalier99} located in the Galactic anti-center region $(l,b)\approx(189^{\circ},+3^{\circ})$ at a distance of about 1.5~kpc in the Gem OB1 association \citep{welsh03}, and is a particularly well-studied SNR.  Figure \ref{figIC443} shows the red image of IC 443 taken during the Second Palomar Observatory Sky Survey.  The remnant is composed of subshells A and B; shell A is to the NE---its center at $\alpha=06^{\rm h}17^{\rm m}08.4^{\rm s}$, $ \delta=+22^{\circ}36'39.4''$ J2000.0 is marked by the cross---while shell B is to the SW.  Adopting a distance of 1.5~kpc, the radii of subshells A and B are about 7~pc and 11~pc, respectively.  Between the subshells is a darker lane that runs across the remnant from the NW to SE.  This is a molecular cloud which has been mapped in $^{12}$CO emission \citep{cornett77,dickman92,zhang09}, and is known to be in the foreground because it absorbs X-rays emitted by the hot remnant interior \citep{troja06}.  Aside from this quiescent foreground cloud, observations of the $J=1\rightarrow0$ line of $^{12}$CO also show shocked molecular material coincident with IC 443 \citep{denoyer79,huang86,dickman92,wang92}.  These shocked molecular clumps first identified by \citet{denoyer79} and \citet{huang86} in CO have also been observed in several atomic and small molecular species \citep[e.g.][]{white87,burton88,vandishoeck93,white94,snell05}, and are thought to be the result of the expanding SNR interacting with the surrounding ISM.  While many of the shocked clumps are coincident with the quiescent gas, it is unclear whether or not they are part of the foreground cloud (i.e. the back portions of the foreground cloud are beginning to interact with the SNR blast wave), or if the foreground cloud is separated from IC 443.

Chemical analyses performed in various studies of the shocked clumps around IC 443 suggest an enhanced ionization rate due to cosmic rays.  \citet{white94} found a C/CO ratio much higher than in typical dense clouds and concluded that shocks and/or a large flux of cosmic rays must be responsible.  Both \citet{claussen97} and \citet{hewitt06} observed OH (1720 MHz) masers toward some of these clumps.  It is thought that this OH is formed when the free electrons produced during ionization events collide with and excite H$_2$, which in turn emits UV photons that dissociate H$_2$O \citep{wardle02}.  In order to convert nearly all of the H$_2$O into OH, thus generating the large column of OH necessary to produce the observed masers, a high ionization rate due to X-rays and/or cosmic rays is required.
Estimates of the ionization rate due to X-rays \citep{yusefzadeh03} and cosmic rays \citep{hewitt09} near IC 443 are similar (a few times 10$^{-16}$~s$^{-1}$), so it may be that both play a role in generating OH.  However, none of these analyses alone can determine exactly how important cosmic-ray ionization and excitation are to the processes considered.

Recently, many studies of IC 443 have focused on the production of pionic gamma-rays via interactions between hadronic cosmic rays and ambient nucleons.  Gamma-ray observations of IC 443 have been performed by {\it EGRET} \citep{esposito96}, {\it MAGIC} \citep{albert07}, {\it VERITAS} \citep{acciari09}, {\it Fermi} LAT \citep{abdo10}, and {\it AGILE} \citep{tavani10}.  All show gamma-ray emission that appears to be coincident with gas in the vicinity of IC 443, thus supporting an enhanced cosmic ray flux in the region.  Because $\pi^0$ production requires cosmic-ray protons with $E_{kin}>280$~MeV, gamma-ray observations cannot constrain the cosmic-ray flux at lower energies.

To investigate the flux of lower-energy cosmic rays, we study the cosmic-ray ionization of H$_2$, a process dominated by protons with $1~{\rm MeV}\leq E_{kin}\leq 1$~GeV \citep{indriolo09,padovani09}.  The ionization rate of H$_2$, $\zeta_2$, can be inferred from observations of H$_3^+$ assuming a rather simple chemical network.  H$_2$ is first ionized, after which the ion collides with another H$_2$, thus forming H$_3^+$.  Either dissociative recombination with electrons (diffuse clouds) or proton transfer to CO, O, and C (dense clouds) are the primary destruction routes for H$_3^+$ depending on the environment.  In this paper, we present observations searching for absorption lines of H$_3^+$ along sight lines which pass through molecular material near IC 443.  We then use the results of these observations in combination with the simple chemical scheme outlined above to infer the cosmic-ray ionization rate of H$_2$.

\section{OBSERVATIONS}

This project examined 6 target sight lines toward the stars ALS 8828, HD 254577, HD 254755, HD 43582, HD 43703, and HD 43907, all of which are shown in Figure \ref{figIC443} to the immediate left of the labels A--F, respectively.  Target selection was based on various criteria, including $L$-band magnitude, previously detected molecules, and evidence that the background stars were in fact behind the SNR \citep{welsh03,hirschauer09}.  Basic properties of these sight lines are available in \citet{hirschauer09}.  Observations focused primarily on transitions arising from the $(J,K)=(1,1)$ and $(1,0)$ levels of the ground vibrational state of H$_3^+$, the only levels significantly populated at average diffuse cloud temperatures ($T\sim60$~K).  Transitions from higher energy levels (e.g. $(2,1)$ and $(3,3)$) were covered as allowed by the instrument, but absorption at these wavelengths was not expected.

\begin{figure}
\epsscale{1.1}
\plotone{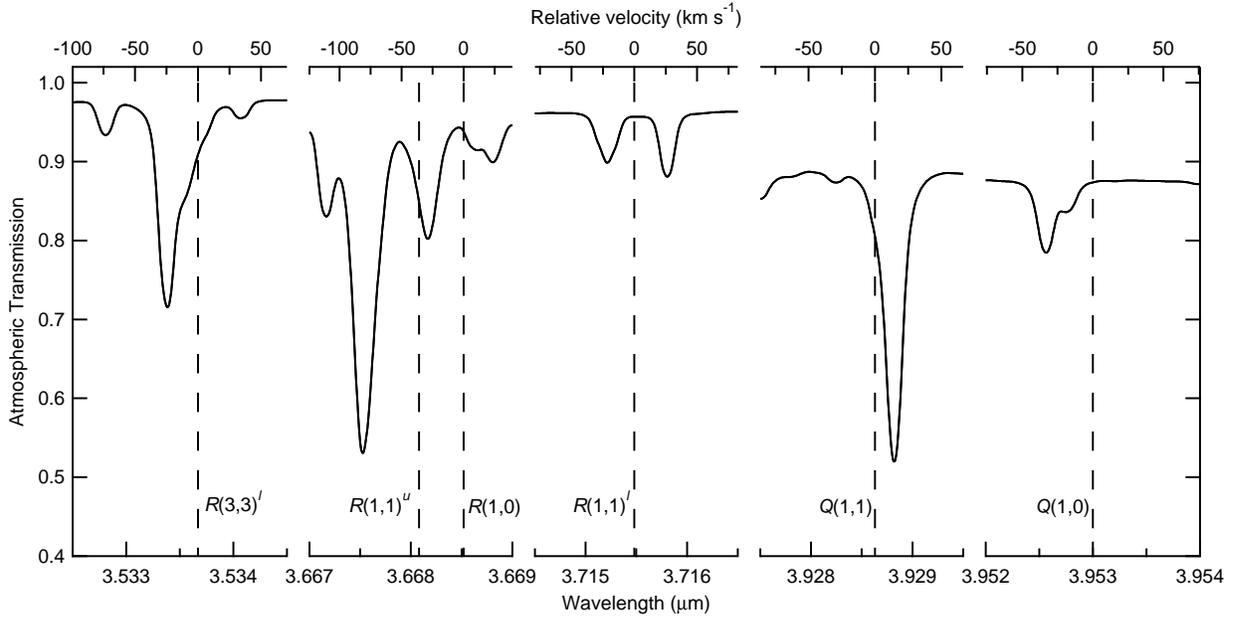}
\caption{Model atmospheric transmission spectra at wavelengths near targeted H$_3^+$ transitions, generated by ATRAN \citep{lord92} assuming observations were performed from Mauna Kea, 1.6~mm precipitable water vapor, and an air mass of 1.15, and smoothed to a resolving power of about 25,000.  Dashed vertical lines mark the rest positions of H$_3^+$ transitions, and are labeled accordingly.  The bottom axes give the rest wavelengths in $\mu$m, while the top axes show the effect of relative motion between the observer and interstellar gas.  Average velocity shifts (due to the Earth's motion and interstellar gas motion) for the spectra taken at Keck and Subaru are about $-18$~km~s$^{-1}$ and $-2$~km~s$^{-1}$, respectively.}
\label{figatm}
\end{figure}

Spectra were obtained using the Near Infrared Echelle Spectrograph \citep[NIRSPEC;][]{mclean98} at the W. M. Keck Observatory, and the Infrared Camera and Spectrograph \citep[IRCS;][]{kobayashi00} at the Subaru Telescope.  All NIRSPEC observations were performed on 2009 Nov 5 and 6 using the 3-pixel (0.432'') slit to provide a resolving power of $\sim25,000$.  The KL filter was used in combination with the echelle grating/cross-disperser settings of 64.82/33.5 in order to simultaneously cover the $R(1,1)^u$, $R(1,0)$, and $R(3,3)^l$ transitions.  All IRCS observations were performed on 2009 Dec 12 and 13 in echelle mode using the 2-pixel (0.14'') slit to provide a resolving power of $\sim17,300$.  The adaptive optics system (AO188) was utilized in order to maximize starlight passing through the narrow slit.  The L filter was used in combination with the echelle grating/cross-disperser settings of 8350/6100 in order to simultaneously cover the $R(1,1)^u$, $R(1,0)$, $R(1,1)^l$, $Q(1,1)$, and $Q(1,0)$ transitions.  Select properties of the six targeted transitions are listed in Table \ref{tbltrans}, and their locations with respect to atmospheric absorption features are shown in Figure \ref{figatm}.  A log containing the list of science targets and exposure times for each night is shown in Table \ref{tblobs}.  In addition to the science targets, the bright, early-type stars $\beta$~Tau and $\zeta$~Tau were observed for use as telluric standards.  These standards were observed so that they would have good air mass matches with the science targets, and to account for atmospheric variability over the course of each night.  For all observations, the star was nodded along the slit in an ABBA pattern in order to facilitate the removal of atmospheric emission lines and dark current via the subtraction of neighboring images.

\section{DATA REDUCTION}

Our data reduction process combines the use of standard IRAF\footnote{http://iraf.noao.edu/} procedures and macros written in IGOR Pro\footnote{http://www.wavemetrics.com/}.  Due to differences between the NIRSPEC and IRCS observations, each data set required slightly different reduction techniques.

\subsection{Keck Data Reduction Process}

A bad pixel map was created from the average of several dark frames, and these pixels were interpolated over in the flat-field and object images.  Images were then cut into two sections, with each section containing one of the orders of interest from the cross-dispersed spectrograph.  These sections were then treated as individual images for the remainder of the reduction process.  The flats were then combined, and each object frame was divided by the normalized, averaged flat-field.  Neighboring AB image pairs were subtracted from each other to remove atmospheric emission and dark current.  One-dimensional spectra were extracted for each order using {\it apall}, and imported to IGOR Pro.

\subsection{Subaru Data Reduction Process}

While no dark frames were taken at Subaru, calibration frames with the lamp off were used to create a bad pixel map.  These pixels were interpolated over in the object and flat-field frames, and again each image was cut into two sections containing an order of interest.  Inspection of the averaged, normalized flat-field frame for each order showed a low signal-to-noise ratio (S/N), so object frames were not divided by these flats.  As before, one-dimensional spectra were extracted with {\it apall} and imported to IGOR Pro. In order to remove a saw-tooth pattern (the result of different readout channels) from these spectra, a moving average was taken for both the odd and even numbered pixels, and the even pixels were then scaled by the ratio of these averages.

\subsection{Shared Reduction Processes}

\begin{figure}
\epsscale{1.2}
\plotone{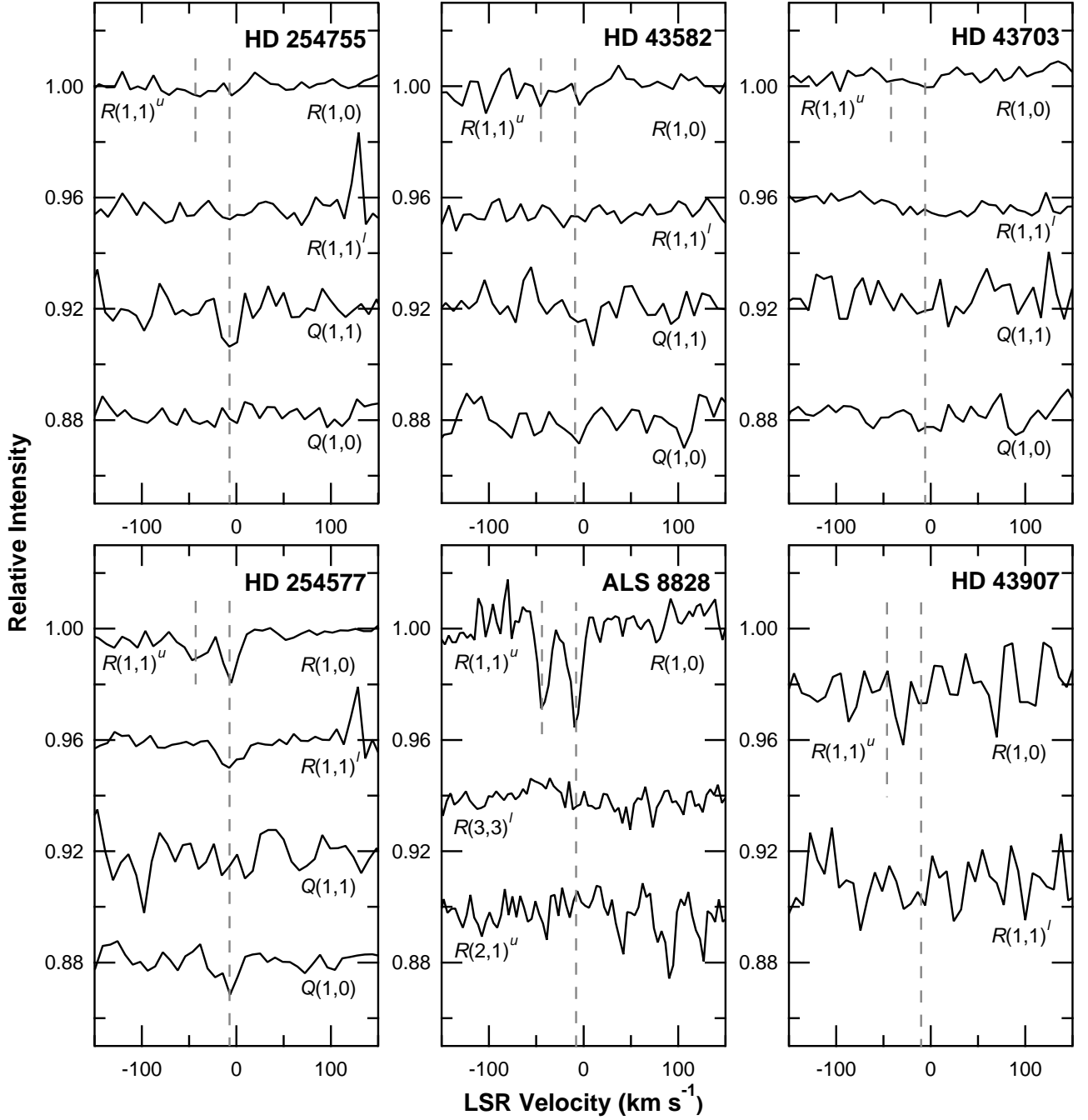}
\caption{Spectra of six stars in the IC 443 region covering various H$_3^+$ transitions.  Vertical dashed lines show the expected position of absorption lines due to H$_3^+$ given the velocities reported by \citet{hirschauer09} for cloud components with the most CH absorption.  The shorter dashed line in the $R(1,1)^u$--$R(1,0)$ spectra shows the position of the $R(1,1)^u$ line, which is 36~km~s$^{-1}$ away from the $R(1,0)$ line.  For HD 254755, HD 43582, HD 43703, and HD 254577, the $R(1,1)^u$--$R(1,0)$ spectra are the combination of NIRSPEC and IRCS data, while the $R(1,1)^l$, $Q(1,1)$, and $Q(1,0)$ spectra are only from IRCS.  All of the ALS 8828 spectra are from NIRSPEC, and HD 43907 spectra are from IRCS.  Of these six sight lines, only HD 254577 and ALS 8828 show H$_3^+$ absorption features.  The $R(1,1)^u$, $R(1,0)$, $R(1,1)^l$, and $Q(1,0)$ lines are visible toward HD 254577 (the lower S/N in the $Q(1,1)$ spectrum and smaller dipole moment of that transition precludes its detection).  For ALS 8828, relatively strong $R(1,1)^u$ and $R(1,0)$ lines are visible.  Even with a large amount of H$_3^+$ along the sight line though, the $R(2,1)^u$ and $R(3,3)^l$ transitions arising from higher energy states are not detected.
}
\label{figspectra}
\end{figure}

Individual spectra within an exposure sequence for a given target were then added together.  In this process, spectra with S/N much lower than average (due to cirrus clouds or a bad nodding sequence) were excluded.  Each summed spectrum was then divided by a telluric standard to remove atmospheric absorption features and normalize the spectrum.  These ratioed spectra were wavelength calibrated with a typical accuracy of $\sim2$~km~s$^{-1}$ using the vacuum wavelengths of the atmospheric absorption lines.  Calibrated spectra were then shifted into the local standard of rest (LSR) frame.  At this point all spectra of a given target from a single telescope were combined using a variance-weighted mean\footnote{This averaging scheme is equivalent to weighting each spectrum by (S/N)$^2$.} ($\bar{x}=\sum_{i=0}^{n} (x_i/\sigma_i^2)/\sum_{i=0}^{n} (1/\sigma_i^2)$, where $\sigma_i$ is the standard deviation on the continuum near the H$_3^+$ line positions)
to produce a final Keck and Subaru spectrum for each sight line.  The $R(1,1)^l$, $Q(1,1)$, and $Q(1,0)$ transitions were only covered at Subaru, while the $R(3,3)^l$ transition was only covered at Keck, so these are the final spectra presented in Figure \ref{figspectra}.  The $R(1,1)^u$ and $R(1,0)$ transitions were covered at both telescopes, so the final Keck and Subaru spectra can be combined to obtain a higher S/N.  The Keck spectra were interpolated onto the lower resolution Subaru wavelength scale, and all spectra were again combined via the weighting scheme described above.  The resulting spectra for our six target sight lines are shown in Figure \ref{figspectra}.

\section{RESULTS}

It is clear from Figure \ref{figspectra} that H$_3^+$ absorption is only detected in the sight lines toward ALS 8828 and HD 254577.  The $R(1,1)^u$ and $R(1,0)$ lines are quite strong toward ALS 8828.  Absorption from the higher energy (3,3) and (2,1) states was not detected, as expected given diffuse molecular cloud conditions.  The sight line toward HD 254577 shows absorption from the $R(1,1)^u$, $R(1,0)$, $R(1,1)^l$ and $Q(1,0)$ transitions of H$_3^+$.  Absorption due to the $Q(1,1)$ transition must also be present, but it is not detected.  This is probably the result of 3 factors: (1) the intrinsic strength of the $Q(1,1)$ transition is the weakest of the 5 transitions examined (see Table \ref{tbltrans}); (2) the spectrum near the $Q(1,1)$ transition has a low S/N due to lower illumination of the echelle order in which it appears; (3) the $Q(1,1)$ transition is overlapped by a strong atmospheric N$_2$O line (see panel 4 of Figure \ref{figatm}), making removal of telluric features uncertain.  Imperfect removal of this atmospheric line is also the most likely cause of the feature in the $Q(1,1)$ spectrum of HD 254755 that appears at the expected velocity.  This feature cannot be due to H$_3^+$, as there is no absorption by any of the other stronger H$_3^+$ transitions which arise from the same state.  The positive spike near 130~km~s$^{-1}$ in the $R(1,1)^l$ spectra of HD 254577 and HD 254755 is an instrumental artifact.  Spectra of HD 43582, HD 43703, and HD 43907 also show no absorption features from H$_3^+$.

\section{ANALYSIS}

Equivalent widths were determined using Gaussian fits to the absorption features.  Uncertainties were determined from the standard deviation, $\sigma$, on the residual continuum after subtracting by the Gaussian line profiles.  Interstellar gas velocities and velocity full width at half-maxima (FWHM) were also determined during this fitting procedure.  In the case of non-detections, upper limits were determined from $3\sigma$ on the continuum across the expected position of a line assuming a FWHM of 16~km~s$^{-1}$ (the resolution obtained with IRCS at Subaru).  Column densities were derived from equivalent widths using the standard relation given optically thin absorption lines and the transition dipole moments and wavelengths listed in Table \ref{tbltrans}.  All of these results are shown in Table \ref{tbllineparam}.

Following the analysis of \citet{indriolo07}, we adopt the simple chemical scheme in diffuse clouds where every ionization of an H$_2$ molecule leads to H$_3^+$, and dissociative recombination with electrons is the dominant mechanism by which H$_3^+$ is destroyed.  This results in the steady-state equation
\begin{equation}
\zeta_2n({\rm H}_2)=k_{e}n_{e}n({\rm H}_3^+),
\label{eqh3plusss}
\end{equation}
\citep{geballe99}, where $\zeta_2$ is the ionization rate of H$_2$, and $k_e$ is the electron recombination rate coefficient of H$_3^+$.  Substituting the electron fraction (defined as $x_e\equiv n_{e}/n_{\rm H}$, where $n_{\rm H}\equiv n({\rm H})+2n({\rm H}_2)$) into equation (\ref{eqh3plusss}) and solving for the ionization rate gives
\begin{equation}
\zeta_2=k_{e}x_{e}n_{\rm H}\frac{n({\rm H}_3^+)}{n({\rm H}_2)}.
\label{eqzeta2numden}
\end{equation}
Although it would be desirable to trace the ionization rate as a function of position throughout the cloud, variations in density along the line of sight cannot be determined via observations.  Instead, we infer the average ionization rate in a cloud by using average number densities.  By definition, $\langle n({\rm H}_3^+)\rangle$ and $\langle n({\rm H}_2)\rangle$ can be replaced with $N({\rm H}_3^+)/L$ and $N({\rm H}_2)/L$, respectively (where $L$ is the cloud path length), thus putting equation (\ref{eqzeta2numden}) in terms of observables. As H$_3^+$ will form wherever there is an appreciable amount of H$_2$, it is reasonable to assume that the path length for both species is the same, such that
\begin{equation}
\zeta_2=k_{e}x_{e}n_{\rm H}\frac{N({\rm H}_3^+)}{N({\rm H}_2)}.
\label{eqzeta2}
\end{equation}
Because the ratio $n({\rm H}_3^+)/n({\rm H}_2)$ is not expected to vary widely in models of diffuse molecular clouds \citep[e.g.][]{neufeld05}, this should give a representative value of the ionization rate throughout the entire cloud.

%All of the variables on the right-hand side of equation (\ref{eqzeta2}) are then determined as follows.

%Solving for the ionization rate, we then have
%\begin{equation}
%\zeta_2=k_{e}n_{e}\frac{N({\rm H}_3^+)}{N({\rm H}_2)}.
%\label{eqzetaderiving}
%\end{equation}
%Finally, we can substitute in the definition for electron fraction ($x_e\equiv n_{e}/n_{\rm H}$, where $n_{\rm H}\equiv n({\rm H})+2n({\rm H}_2)$) to find

Assuming that the vast majority of electrons in diffuse molecular clouds come from photoionized carbon, $x_e$ can be approximated by $N({\rm C}^+)/N_{\rm H}$ ($N_{\rm H}$ is the column density analog to $n_{\rm H}$), which was found to be about $1.5\times10^{-4}$ along multiple diffuse cloud sight lines \citep{cardelli96,sofia04}.  The hydrogen number density can be estimated by both a rotation-excitation analysis of C$_2$ observations and a restricted chemical analysis based on CN observations \citep{hirschauer09}.  The C$_2$ analysis also gives a best-fit kinetic temperature which we use in calculating $k_e$.  The temperature dependency of $k_e$ as determined from laboratory work is reported in \citet{mccall04}.  While molecular hydrogen has not been observed in absorption along any of our target sight lines, abundances of H$_2$ and CH tend to be linearly related in diffuse clouds \citep{federman82,mattila86,sheffer08}.  We use the relationship derived from the largest, most recent data set---$N({\rm CH})/N({\rm H}_2)=3.5_{-1.4}^{+2.1}\times10^{-8}$ \citep{sheffer08}---in combination with CH column densities reported by \citet{hirschauer09} to estimate $N({\rm H}_2)$.  Finally, the total H$_3^+$ column density is determined by adding $N(1,0)$ and $N(1,1)$.  These input values and/or the parameters on which they depend, as well as the inferred ionization rates are shown in Table \ref{tblionization}.

While H$_2$ is ionized by both cosmic rays and X-rays, most of the X-ray flux should be attenuated in a relatively thin layer at the cloud exterior \citep{glassgold74}.  The ionization rate due to X-rays at the edge of IC 443 was estimated to be $\zeta_{\rm X}=3.6\times10^{-16}$~s$^{-1}$ \citep{yusefzadeh03}, and must be much lower in cloud interiors.  As a result, the ionization rates we infer should be primarily due to cosmic rays.

\section{DISCUSSION}

Having computed the cosmic-ray ionization rate for clouds in the vicinity of IC 443, we compare our results to those from previous studies.  The average ionization rate in diffuse molecular clouds found by \citet{indriolo07} using H$_3^+$ was $\zeta_2=4\times10^{-16}$~s$^{-1}$, several times lower than found toward ALS 8828 and HD 254577 ($\zeta_2=16^{+8}_{-12}\times10^{-16}$~s$^{-1}$ and $26^{+13}_{-19}\times10^{-16}$~s$^{-1}$, respectively).  In fact, the ionization rates inferred for these 2 sight lines are more than twice the highest rates previously found in diffuse molecular clouds toward $\zeta$~Per and X~Per ($\zeta_2\approx7\times10^{-16}$~s$^{-1}$).  While ALS 8828 and HD 254577 present exceptionally high ionization rates, the other 3 sight lines observed near IC 443 do not (due to the low S/N obtained toward HD 43907, the derived upper limit for that particular sight line is not exceptionally meaningful, and so we exclude it from further consideration).  Instead, the $3\sigma$ upper limits for $\zeta_2$ presented in Table \ref{tblionization} are consistent with ionization rates of a few times $10^{-16}$~s$^{-1}$, typical of diffuse molecular clouds.  These differences are quite striking, and warrant discussion.

There are two plausible explanations for why H$_3^+$ would be detected toward ALS 8828 and HD 254577 but not HD 254755, HD 43582, HD 43703, and HD 43907, and they can most easily be seen when equation (\ref{eqzeta2}) is rearranged to show that $N({\rm H}_3^+) \propto \zeta_2/(x_{e}n_{\rm H})$.  Given this scaling we can posit that either the product $x_{e}n_{\rm H}$ (i.e. the electron density) is lower along these 2 sight lines, or $\zeta_2$ is higher, and we examine these possibilities in turn.

%by rearranging equation (\ref{eqzeta2}) to solve for $N({\rm H}_3^+)$.  In this form, it is apparent that the H$_3^+$ column density is proportional to the cosmic-ray ionization rate, and inversely related to the hydrogen density and electron fraction.  From these relations,

\subsection{Lower Electron Density}

As stated in Section 5, we have assumed an electron fraction that is consistent with observations of C$^+$ in several diffuse molecular clouds.  In denser environments though, the predominant form of carbon shifts from C$^+$ to C, and eventually to CO, thus decreasing the electron density.  Adopting a reduced value for $x_{e}n_{\rm H}$ requires a corresponding decrease in $\zeta_2$ to match the observed H$_3^+$ column density.  It could then be argued that the enhanced ionization rate we calculate for the 2 sight lines where we detect H$_3^+$ is actually just an artifact of not recognizing a decreased destruction rate.

However, there are observations which seem to argue against this possibility.  The C$_2$ rotation-excitation and CN restricted chemical analyses performed by \citet{hirschauer09} suggest densities of $200-400$~cm$^{-3}$, typical of diffuse molecular clouds, not dense clouds.  Also, we can estimate the fractional abundance of CO ($x({\rm CO})=N({\rm CO})/N_{\rm H})$ in the observed sight lines and compare it to the solar system abundance of carbon \citep[$x({\rm C_{tot}})=2.9\times10^{-4}$;][]{lodders03} to determine if CO is the dominant carbon bearing species.  We estimate $N_{\rm H}$ from the color excess (see Table \ref{tblionization}), and use observed relationships between $N({\rm CH})$, $N({\rm CN})$, and $N({\rm CO})$ \citep{sonnentrucker07,sheffer08}, in concert with CH and CN column densities \citep{hirschauer09} to estimate $N({\rm CO})$.  In the ALS 8828 and HD 254577 sight lines $x({\rm CO})\sim5\times10^{-6}$ and $2\times10^{-6}$, respectively, much smaller than the assumed total carbon budget.  In the other 3 sight lines $x({\rm CO})$ ranges from about $1\times10^{-7}$ to about $1\times10^{-6}$.  These estimates show that most carbon is not in the form of CO, but does not rule out C as the dominant carbon bearing species.  To do so, we use the observed relationship between CO/H$_2$ and (C+CO)/C$_{\rm tot}$ shown in Figure 6 of \citet{burgh10}.  For CO/H$_2\sim8\times10^{-6}$---the largest value estimated along any of our sight lines---observations show that both CO and C account for only a small fraction of the total carbon budget, thus indicating that carbon is predominantly in ionized form.

To improve upon these rough arguments though, observations yielding the relative abundances of C$^+$, C, and CO are necessary.  The $v$=1--0 fundamental and $v$=2--0 overtone rovibrational bands of CO near 4.6~$\mu$m and 2.3~$\mu$m, respectively, can be observed with NIRSPEC and IRCS.  Various electronic transitions of CO and C~\textsc{i} are available in the far ultraviolet (1100~\AA--1700~\AA), and can be observed with either COS or STIS aboard {\it Hubble}.  Finally, a weak intersystem line of C~\textsc{ii} is at 2325~\AA, and may also be observable with COS and/or STIS.  Combined, these observations would allow us to determine the predominant carbon-bearing species along each sight line, and give us a better understanding of cloud conditions being probed.

\subsection{Higher Ionization Rate}

If the gas conditions in all of our observed sight lines are similar, then the cosmic-ray ionization rate must be higher toward ALS 8828 and HD 254577.  Such varied ionization rates can be the result of differing cosmic-ray fluxes in each sight line.  If we assume that the SNR accelerates particles isotropically (i.e. the spectrum of cosmic rays leaving the remnant is identical everywhere along the blast wave) then the different cosmic-ray spectra operating in each sight line must be due to propagation effects.

To determine whether or not cosmic rays accelerated by IC 443 can even produce the high inferred ionization rates, we use the methods described in \citet{indriolo09} to compute the expected ionization rate for various cosmic-ray spectra.  \citet{abdo10}, \citet{torres08}, and \citet{torres10} constrain the proton spectrum above $\sim100$~MeV near IC 443 from the observed gamma-ray spectrum.  Although the broken power-law proton spectrum in \citet{abdo10} is given as a power law in kinetic energy (${\rm flux}\propto E_{kin}^{-2.09}$ when $E_{kin}<69$~GeV), we change this to a power law in momentum (${\rm flux}\propto p^{-2.09}$, where $pc=[(E_{kin}+m_pc^2)^2-(m_pc^2)^2]^{0.5}$) to account for the fact that diffusive shock acceleration is expected to produce a spectrum of this form.
%particles distributed as a power law in momentum.
This substitution only differs from the relation considered by \citet{abdo10} in the non-relativistic regime where $p\propto E_{kin}^{0.5}$.  As we extrapolate the spectrum to lower energies then, the particle distribution is flattened relative to a pure power law in kinetic energy.

%and thus the spectrum is flattened relative to a pure power law in kinetic energy.
%we extrapolate it to lower energies as a power law in momentum using the relation $pc=[(E_{kin}+m_pc^2)^2-(m_pc^2)^2]^{0.5}$, where $m_pc^2$ is the proton rest-mass energy.  This accounts for the fact that diffusive shock acceleration is expected to produce particles distributed as a power law in momentum, and flattens the cosmic-ray spectrum as particle energies change from the relativistic ($p\propto E_{kin}$) to non-relativistic ($p\propto E_{kin}^{0.5}$) regime.

Integrating this extrapolated spectrum to a low-energy cutoff of 5~MeV, we find $\zeta_2\sim10^{-14}$~s$^{-1}$.  A 5~MeV cutoff was used because particles of this energy have a range of a few times $10^{21}$~cm$^{-2}$ \citep{padovani09}, similar to the sight lines we consider here.  Using the $d=10$~pc continuous injection spectrum (also extrapolated to low energies as above) from \citet{torres08} produces similar results, while their $d=30$~pc spectrum fails to reproduce even the ionization rate predicted by the local interstellar cosmic-ray spectrum \citep[$\zeta_2\sim4\times10^{-17}$~s$^{-1}$;][]{webber98}.  Note that the difference between the 10~pc and 30~pc theoretical spectra is {\it not} the result of energy losses, but due to the fact that lower energy particles have not yet had sufficient time to travel far from IC 443 given its age of 30,000 yr.  Although these spectra are not particularly well suited for estimating the cosmic-ray ionization rate---they are based on observations which depend on processes requiring $E_{kin}>280$~MeV and so are not well constrained at energies of a few MeV where ionization is much more efficient---they do suggest that cosmic-rays accelerated by IC 443 are capable of generating the ionization rate inferred from H$_3^+$, and also provide independent constraints on the flux of high-energy cosmic rays near IC 443 which complement the low-energy component studied in this paper.

The propagation effects included in the model cosmic-ray spectra presented in \citet{torres08,torres10} may also be able to explain the differences inferred in $\zeta_2$.  Those authors suggest that the difference in centroid position between the GeV ({\it EGRET} and {\it Fermi} LAT) and TeV ({\it MAGIC} and {\it VERITAS}) gamma-ray sources can be explained by having the gamma rays of different energies originate in separate clouds.  They propose that the lower energy gamma rays arise from $\pi^0$ decay in a cloud 3--6~pc away from the expanding shell of IC 443, and the higher energy gamma rays in a cloud about 10~pc in front of the SNR.  This explanation requires that cosmic-ray propagation is energy dependent, such that high energy particles have diffused farther away from the SNR than low energy particles.  In such a model, the cosmic-ray spectrum varies as a function of position, and so the ionization rate must as well.

%, the former 3--6~pc away from the expanding shell of IC 443, and the latter about 10~pc in front of the SNR.  These correspond to the shocked material interacting with the remnant and the foreground quiescent cloud, respectively.
%This picture is supported by the recent gamma-ray observations where the lower energy centroids ({\it EGRET} and {\it Fermi} LAT) are consistent with each other, but removed from the higher energy centroids ({\it MAGIC} and {\it VERITAS}), which are also consistent with each other \citep[see Figure 5 from][]{abdo10}.
%Also, all of these gamma-ray observations provide independent measures of the high-energy cosmic-ray flux near IC 443 which compliment the low-energy tracers we study.

Because lower energy cosmic rays have yet to propagate very far from IC 443, the ionization rate should decrease with increased distance away from the SNR.  If the clouds probed by ALS 8828 and HD 254577 are closer to IC 443 than the clouds probed by the other 3 sight lines, then the difference in inferred ionization rates is easily explained.  The positions of our target sight lines with respect to IC 443 are shown in Figure \ref{figIC443}, and the on-sky distances from the center of subshell A to each of the background stars is listed in Table \ref{tblionization}.  Of the 5 sight lines, only HD 43703 is a considerable distance away from the remnant, so differences in the remaining 4 sight lines must be due to line-of-sight distances.  Gas velocities for the dominant CH components reported by \citet{hirschauer09} vary by only about 3~km~s$^{-1}$ between all of our sight lines, suggesting that the absorption may arise from the same cloud complex, but because IC 443 is located near the Galactic anti-center such an analysis is highly uncertain.  The HD 254577 sight line passes through regions of HCO$^+$ emission \citep{dickman92} and H$_2$ emission \citep{burton88,inoue93,rho01}, both of which trace shocked gas, and is in close proximity to an OH (1720~MHz) maser which requires shocked gas and a high ionization rate \citep{hewitt06}, so it is plausible that the observed H$_3^+$ absorption arises in material very close to the SNR shock.
%In addition, the velocity of this maser, $-6.85$~km~s$^{-1}$, is consistent with the H$_3^+$ velocities reported in Table \ref{tbllineparam}.
The ALS 8828 sight line, however, is not coincident with shock tracers, so it is unclear at this location how close the foreground cloud is to the SNR.  Still, given the drastic difference in the 10~pc and 30~pc cosmic-ray spectra from \citet{torres08} (see their Figure 1), the gas probed by the 3 sight lines without observed H$_3^+$ would not have to be that much farther away than the gas probed by ALS 8828 to explain the inferred ionization rates; something on the order of 10~pc farther away would suffice.

Aside from the distance between the site of particle acceleration and the clouds in question, various other propagation and acceleration effects could account for the difference in inferred ionization rates.  Cosmic rays diffuse through space as they scatter off of Alfv\'{e}n waves which are presumed to be generated by the particles themselves.  For clouds with higher densities of neutral gas, the damping of these waves (via ion-neutral collisions) becomes more efficient and the streaming velocity of cosmic rays thus increases \citep{padoan05}.  Instead of diffusing then, particles will free-stream and spend much less time in the cloud (i.e. have fewer chances to ionize ambient material).  As a result, regions of low gas density should be expected to have higher ionization rates than regions of high gas density.  Another possibility is that the net flux of cosmic rays into a cloud (due to ionization losses, nuclear interactions, etc. within the cloud) sets up an anisotropy that causes the growth of Alfv\'{e}n waves in the plasma surrounding the cloud.  Lower energy particles ($E_{kin}$ less than a few hundred MeV) scatter off of these waves and are impeded from entering the cloud \citep{skilling76}.  Because the particles most efficient at ionizing hydrogen are excluded from denser clouds, this effect also predicts a higher ionization rate in regions of lower density.  However, due to the similar densities reported in \citet{hirschauer09} for our target sight lines, these effects seem unlikely candidates for causing the difference in inferred ionization rates.

%in the low density region surrounding the cloud which generates Alfv\'{e}n waves which scatter lower energy particles ($E_{kin}$ less than a few hundred MeV) and exclude them from entering the cloud \citep{skilling76}.

The final effect we consider in attempting to explain these variations in the ionization rate is the escape of cosmic rays upstream from the shock where diffusive shock acceleration occurs (i.e. away from the SNR).  This subject has been the focus of several recent studies \citep[e.g.][]{caprioli09,caprioli10,reville09,ohira10} which find that particles {\it can} escape in the upstream direction, although these tend to be only the particles with the highest energies.  For the discussion above, we have assumed that low-energy cosmic rays have escaped from the shock and are diffusing away from the SNR.  However, if low-energy cosmic rays do not escape, but are instead preferentially advected downstream (i.e. into the SNR), then the ionization rate in the post-shock gas inside the SNR should be higher than in the gas exterior to the remnant.  The differing ionization rates could then be explained if the sight lines toward ALS 8828 and HD 254577 probed gas interior to IC 443.  As mentioned above, the HD 254577 sight line is coincident with various shock tracers, as well as an OH (1720~MHz) maser (which arises from the post-shock gas inside the SNR). Additionally, the velocity of this maser, $-6.85$~km~s$^{-1}$, is consistent with the H$_3^+$ velocities reported in Table \ref{tbllineparam}, making it highly plausible that the H$_3^+$ absorption toward HD 254577 arises from shocked gas inside of IC 443.  Consequently, the inability of low-energy cosmic rays to escape from IC 443 provides an alternative to the diffusion of particles and differing distances between the remnant and gas probed by our sight lines in explaining the inferred ionization rates.

\subsection{Implications}

Given either of the cases discussed above (low electron density or high ionization rate), we can comment on the flux of low-energy cosmic rays accelerated by SNRs.  In the case that the exceptional H$_3^+$ column densities observed are due to a lower destruction rate (i.e. lower electron density), then the ionization rate near IC 443 is no higher than already found toward various diffuse molecular cloud sight lines.  This would indicate that the flux of low energy cosmic rays near SNRs is not substantially different than in the Galactic ISM, and suggest that either SNRs are not the primary accelerators of such particles, or that low energy particles have yet to escape from IC 443.

In the case that the inferred ionization rates of a few times $10^{-15}$~s$^{-1}$ are correct, IC 443 must be accelerating a large population of low-energy cosmic rays.  Either this population must be escaping upstream from the site of diffusive shock acceleration (i.e. traveling outward from the SNR shock) such that the clouds closest to the remnant are experiencing a large flux of cosmic rays, or the 2 sight lines with H$_3^+$ detections probe gas inside of IC 443 where low-energy cosmic rays have been advected downstream.  In either situation, it is unclear if such a population of cosmic rays accelerated by all SNRs within the Galaxy will propagate far enough from their sources to affect the flux of cosmic rays at some arbitrary position.  As a result, it is difficult to definitively say whether or not SNRs are responsible for accelerating the large flux of low-energy cosmic rays necessary to produce the $\zeta_2\sim4\times10^{-16}$~s$^{-1}$ ionization rate inferred in many diffuse Galactic sight lines.

\section{CONCLUSIONS}

We have searched for H$_3^+$ absorption along 6 sight lines that pass through molecular material in the vicinity of the SNR IC 443.  Two of the observed sight lines, ALS 8828 and HD 254577, have large column densities of H$_3^+$, while the other 4 show no absorption features.  The cosmic-ray ionization rates inferred from the 2 detections are a few times $10^{-15}$~s$^{-1}$, higher than ever previously found in diffuse molecular clouds.  Upper limits to the ionization rate in the other sight lines, however, are consistent with values found along average Galactic sight lines of about $4\times10^{-16}$~s$^{-1}$.  These differences may be due to overestimates of the electron fraction or cosmic-ray propagation and acceleration effects, but the complexity of the region makes it difficult to attribute the results to any one cause.  Future observations of C~\textsc{ii}, C~\textsc{i}, and CO toward our target sight lines should allow us to better discriminate between the two possibilities, and thus determine whether or not IC 443 produces a large flux of low-energy cosmic rays.

In addition, surveys of H$_3^+$ near IC 443 and other SNRs thought to be interacting with molecular clouds (e.g. Vela, W 28, W 44, W 51C) should allow us to further investigate cosmic-ray acceleration in such environments.  By more extensively mapping H$_3^+$ absorption near supernova remnants, we can determine where the H$_3^+$ resides (interior post-shock gas or exterior pre-shock gas), and thus where the flux of low-energy cosmic rays is highest.  Such observations may also provide insight into the efficiency with which accelerated particles are advected downstream into remnants, and so add important constraints to models of cosmic-ray acceleration.

%Still, this would only put constraints on cosmic-ray acceleration by IC 443.  Surveys of H$_3^+$ near more SNRs known to be interacting with molecular clouds (e.g. Vela, W 28, W 44, W 51C), as well as the necessary complimentary observations, would be highly advantageous in advancing our understanding of the cosmic ray flux in such regions.

\mbox{}

The authors would like to thank Steve Federman, Farhad Yusef-Zadeh and the anonymous referee for helpful comments and suggestions.  N. I. and B. J. M. are supported by NSF grant PHY 08-55633.  G. A. B. is supported by NSF grant AST 07-08922.  T. O. is supported by NSF grant AST 08-49577.  T. R. G.'s research is supported by the Gemini Observatory, which is operated by the Association of Universities for Research in Astronomy, Inc., on behalf of the international Gemini partnership of Argentina, Australia, Brazil, Canada, Chile, the United Kingdom, and the United States of America.  The work of B. D. F. was partially supported by the NASA Astrophysics Theory Program through award NNX10AC86G.  The Digitized Sky Surveys were produced at the Space Telescope Science Institute under U.S. Government grant NAG W-2166. The images of these surveys are based on photographic data obtained using the Oschin Schmidt Telescope on Palomar Mountain and the UK Schmidt Telescope. The plates were processed into the present compressed digital form with the permission of these institutions.  The Second Palomar Observatory Sky Survey (POSS-II) was made by the California Institute of Technology with funds from the National Science Foundation, the National Geographic Society, the Sloan Foundation, the Samuel Oschin Foundation, and the Eastman Kodak Corporation.

%%%%%%%%%%%%%%%%%%%%%%%figures%%%%%%%%%%%%%%%%%%%%%%%%%%%%%%%%%%%%%%%%%%%

\clearpage

\clearpage

%%%%%%%%%%%%%%%%%%%%%%%%%%%%%%%%%%%%%%%%%%%%%%%%%%%%%%%%%%%%%%%%%%%%%%%%%%

\clearpage

%%%%%%%%%%%%%%%%%%%%%%%%%%%%%%%%%%%%%%%%%%%%%%%%%%%%%%%%%%%%%%%%%%%%%%%%%%%%

%%%%%%%%%%%%%%%%%%%%%%%%tables%%%%%%%%%%%%%%%%%%%%%%%%%%%%%%%%%%%%%%%%%%%%

\clearpage

\begin{deluxetable}{lcc}
\tablecaption{H$_3^+$ Transition Properties \label{tbltrans}}
\tablehead{
 & \colhead{Wavelength} & \colhead{$|\mu|^2$} \\
\colhead{Transition} & \colhead{($\mu$m)} & \colhead{(D$^2$)}
}
\startdata
$R(1,1)^u$ & 3.668083 & 0.0158 \\
$R(1,0)$   & 3.668516 & 0.0259 \\
$R(1,1)^l$ & 3.715479 & 0.0141 \\
$Q(1,1)$   & 3.928625 & 0.0128 \\
$Q(1,0)$   & 3.953000 & 0.0254 \\
$R(3,3)^l$ & 3.533666 & 0.0191
\enddata
\tablecomments{Wavelengths and dipole moments for targeted transitions in the $\nu_2\leftarrow0$ band of H$_3^+$ \citep[values from][and references therein]{goto02}.}
\end{deluxetable}

%%%%%%%%%%%%%%%%%%%%%%%%%%%%%%%%%%%%%%%%%%%%%%%%%%%%%%%%%%%%%%%%%%%%%%%%%

\clearpage

\begin{deluxetable}{llcc}
\tablecaption{Observations \label{tblobs}}
\tablehead{
 & & & \colhead{Integration Time} \\
\colhead{Object} & \colhead{Date(s) of Observation} & \colhead{Telescope} & \colhead{(min)}
}

\startdata
ALS 8828  & 2009 Nov 5  & Keck   & 40  \\
          & 2009 Nov 6  & Keck   & 24  \\
HD 254577 & 2009 Nov 5  & Keck   & 40  \\
          & 2009 Nov 6  & Keck   & 20  \\
          & 2009 Dec 12 & Subaru & 96  \\
          & 2009 Dec 13 & Subaru & 36  \\
HD 254755 & 2009 Nov 5  & Keck   & 40  \\
          & 2009 Nov 6  & Keck   & 20  \\
          & 2009 Dec 13 & Subaru & 120 \\
HD 43582  & 2009 Nov 5  & Keck   & 28  \\
          & 2009 Nov 6  & Keck   & 40  \\
          & 2009 Dec 12 & Subaru & 120 \\
HD 43703  & 2009 Nov 6  & Keck   & 68  \\
          & 2009 Dec 13 & Subaru & 120 \\
HD 43907  & 2009 Dec 12 & Subaru & 42
\enddata

\end{deluxetable}

%%%%%%%%%%%%%%%%%%%%%%%%%%%%%%%%%%%%%%%%%%%%%%%%%%%%%%%%%%

\clearpage
\begin{deluxetable}{llcccccc}
\tablecaption{Absorption Line Parameters \label{tbllineparam}}
\tablehead{
 & & \colhead{$v_{\rm LSR}$} & \colhead{FWHM} & \colhead{$W_{\lambda}$} & \colhead{$\sigma(W_{\lambda})$} &
\colhead{$N(J,K)$} & \colhead{$\sigma(N)$} \\
\colhead{Object} & \colhead{Transition} & \colhead{(km s$^{-1}$)} & \colhead{(km s$^{-1}$)} &
\colhead{($10^{-6}~\mu$m)} & \colhead{($10^{-6}~\mu$m)} & \colhead{($10^{14}$ cm$^{-2}$)} & \colhead{($10^{14}$ cm$^{-2}$)}
}

\startdata
ALS 8828  & $R(1,1)^u$ & -6.0 & 14.0 & 6.4    & 0.5 & 2.64    & 0.22 \\
          & $R(1,0)$   & -8.3 & 14.1 & 7.0    & 0.5 & 1.76    & 0.14 \\
HD 254577 & $R(1,1)^u$ & -6.7 & 24.9 & 3.0    & 0.4 & 1.22    & 0.15 \\
          & $R(1,0)$   & -6.2 & 16.2 & 3.8    & 0.3 & 0.95    & 0.07 \\
          & $R(1,1)^l$ & -6.2 & 27.0 & 3.4    & 0.4 & 1.55    & 0.17 \\
          & $Q(1,1)$   &  ... &   16 & $<3.2$ & ... & $<1.54$ & ... \\
          & $Q(1,0)$   & -6.1 & 14.3 & 2.5    & 0.5 & 0.59    & 0.12 \\
HD 254755 & $R(1,1)^u$ &  ... &   16 & $<1.2$ & ... & $<0.49$ & ... \\
          & $R(1,0)$   &  ... &   16 & $<1.2$ & ... & $<0.30$ & ... \\
          & $R(1,1)^l$ &  ... &   16 & $<1.4$ & ... & $<0.64$ & ... \\
          & $Q(1,1)$   &  ... &   16 & $<3.3$ & ... & $<1.58$ & ... \\
          & $Q(1,0)$   &  ... &   16 & $<1.5$ & ... & $<0.36$ & ... \\
HD 43582  & $R(1,1)^u$ &  ... &   16 & $<1.9$ & ... & $<0.78$ & ... \\
          & $R(1,0)$   &  ... &   16 & $<1.9$ & ... & $<0.48$ & ... \\
          & $R(1,1)^l$ &  ... &   16 & $<1.5$ & ... & $<0.68$ & ... \\
          & $Q(1,1)$   &  ... &   16 & $<3.2$ & ... & $<1.55$ & ... \\
          & $Q(1,0)$   &  ... &   16 & $<3.0$ & ... & $<0.71$ & ... \\
HD 43703  & $R(1,1)^u$ &  ... &   16 & $<1.2$ & ... & $<0.50$ & ... \\
          & $R(1,0)$   &  ... &   16 & $<1.2$ & ... & $<0.31$ & ... \\
          & $R(1,1)^l$ &  ... &   16 & $<1.3$ & ... & $<0.62$ & ... \\
          & $Q(1,1)$   &  ... &   16 & $<3.5$ & ... & $<1.68$ & ... \\
          & $Q(1,0)$   &  ... &   16 & $<2.3$ & ... & $<0.56$ & ... \\
HD 43907  & $R(1,1)^u$ &  ... &   16 & $<4.4$ & ... & $<1.83$ & ... \\
          & $R(1,0)$   &  ... &   16 & $<4.4$ & ... & $<1.12$ & ... \\
          & $R(1,1)^l$ &  ... &   16 & $<5.3$ & ... & $<2.41$ & ...
\enddata
\tablecomments{Column 3 ($v_{\rm LSR}$) gives the interstellar gas velocity in the local standard of rest frame.  Column 4 (FWHM) gives the full width at half-maximum of the absorption features.  In the case of non-detections, the FWHM was set to 16~km~s$^{-1}$, the resolving power of IRCS on Subaru in our particular setup, for the purpose of computing column density upper limits.  Columns 5 and 6 show the equivalent width, $W_{\lambda}$, and its $1\sigma$ uncertainty, $\sigma(W_{\lambda})$, respectively.  Upper limits to $W_{\lambda}$ are equal to $3\sigma(W_{\lambda})$.  Columns 7 and 8 give the column density of H$_3^+$ in the state each transition probes, $N(J,K)$, and its uncertainty, $\sigma(N)$, respectively.  Upper limits to the H$_3^+$ column density are equal to $3\sigma(N)$.
}

\end{deluxetable}

%%%%%%%%%%%%%%%%%%%%%%%%%%%%%%%%%%%%%%%%%%%%%%%%%%%%%%%%%%%%%%%%%%%%%%%%

\clearpage

\begin{deluxetable}{lcccccccc}
\tablecaption{Target Sight Line Properties \label{tblionization}}
\tablehead{
 & \colhead{$r$} & \colhead{$T$} & \colhead{$n_{\rm H}$} & \colhead{$N_{\rm H}$} & \colhead{$L$} & \colhead{$N({\rm H}_2)$} &  \colhead{$N({\rm H_3^+})$} & \colhead{$\zeta_2$} \\
\colhead{Target} & \colhead{(pc)} & \colhead{(K)} & \colhead{(cm$^{-3}$)} & \colhead{($10^{21}~{\rm cm}^{-2}$)} & \colhead{(pc)} & \colhead{($10^{21}~{\rm cm}^{-2}$)} &   \colhead{($10^{14}~{\rm cm}^{-2}$)} & \colhead{($10^{-16}$~s$^{-1}$)}
}
\startdata
ALS 8828  & 6.8\tablenotemark{a}  & 60\tablenotemark{b} & 300\tablenotemark{c} & 3.0 & 3.2 & $2.1^{+1.4}_{-0.8}$ & $4.4\pm0.26$ & $16^{+8}_{-12}$ \\
HD 254577 & 7.0\tablenotemark{a}  & 35\tablenotemark{d} & 325\tablenotemark{d} & 3.6 & 3.6 & $0.9^{+0.6}_{-0.3}$ & $2.2\pm0.34$ & $26^{+13}_{-19}$ \\
HD 254755 & 8.6                   & 35\tablenotemark{d} & 200\tablenotemark{d} & 2.5 & 4.0 & $1.1^{+0.7}_{-0.4}$ & $<0.6$       & $<3.5$ \\
HD 43582  & 5.4\tablenotemark{a}  & 60\tablenotemark{b} & 200\tablenotemark{e} & 1.9 & 3.0 & $0.5^{+0.3}_{-0.2}$ & $<0.8$       & $<9.0$ \\
HD 43703  & 13.9                  & 60\tablenotemark{b} & 300\tablenotemark{c} & 1.7 & 1.8 & $0.8^{+0.5}_{-0.3}$ & $<0.6$       & $<5.7$ \\
HD 43907  & 20.6                  & 60\tablenotemark{b} & 300\tablenotemark{c} & 1.5 & 1.6 & $0.4^{+0.3}_{-0.2}$ & $<2.1$       & $<40$
\enddata
\tablecomments{Various parameters used in our analysis for the target sight lines in this study.  Column 2 gives the on-sky distance ($r$) from the center of shell A of IC 443 ($\alpha=06^{\rm h}17^{\rm m}08.4^{\rm s}$, $ \delta=+22^{\circ}36'39.4''$ J2000.0) to each sight line assuming the remnant is at a distance of 1.5~kpc.  The radius of shell A is about 7~pc.  Temperatures ($T$) were taken from the C$_2$ rotation-excitation analysis when available or set to 60~K, and number densities ($n_{\rm H}$) were taken from either the C$_2$ rotation-excitation analysis or restricted chemical analysis, both reported in \citet{hirschauer09}.  Uncertainties in $n_{\rm H}$ are taken to be $\pm100$~cm$^{-3}$.  Color excesses were decreased by 0.3~mag to remove the contribution from foreground gas \citep[color excesses and the foreground correction are given in][]{hirschauer09}, and the relationship $N_{\rm H}\approx E(B-V)\times5.8\times10^{21}$ cm$^{-2}$ mag$^{-1}$ \citep{bohlin78,rachford02} was used to compute the total hydrogen column densities, $N_{\rm H}$.  Path lengths ($L$) were calculated from $n_{\rm H}$ and $N_{\rm H}$.  Molecular hydrogen column densities were calculated from the relationship $N({\rm CH})/N({\rm H}_2)=3.5_{-1.4}^{+2.1}\times10^{-8}$ \citep{sheffer08} using the dominant CH components from \citet{hirschauer09}.  Uncertainties in $N({\rm H}_2)$ are dominated by the scatter in the above relationship, not uncertainties in $N({\rm CH})$.  Upper limits for $N({\rm H}_3^+)$ are the $3\sigma$ uncertainties from the observations, and the upper limits for $\zeta_2$ are based solely on those values.  To account for the uncertainty in $N({\rm H}_2)$, the upper limits for $\zeta_2$ should be multiplied by 1.5.}
\tablenotetext{a}{This sight lines passes through the remnant.}
\tablenotetext{b}{Kinetic temperature assumed in restricted chemical analysis.}
\tablenotetext{c}{Average density from restricted chemical analysis.}
\tablenotetext{d}{Determined from C$_2$ rotation-excitation analysis.}
\tablenotetext{e}{As no CN was detected toward HD 43582, here we have taken the lower bound on the density from the restricted chemical analysis for other sight lines.}
\end{deluxetable}

%ALS 8828  & 4.8 & 2.1 &  &  &  & 4.4    & 0.26 &  \\
%HD 254577 & 5.3 & 0.9 & 200--450 & 10--60 &  & 2.2    & 0.13 &  \\
%HD 254755 & 4.2 & 1.1 & 200      & 30--40 &  & $<0.6$ & ...  &  \\
%HD 43582  & 3.6 & 0.5 &  &  &  & $<0.8$ & ...  &  \\
%HD 43703  & 3.4 & 0.8 &  &  &  & $<0.6$ & ...  &  \\
%HD 43907  & 3.2 & 0.4 &  &  &  & $<2.1$ & ...  &

%E(B-V)& N(CH)   & vLSR
%0.82 & $72\pm3$ & -7.9 \\
%0.92 & $32\pm1$ & -7.1 \\
%0.73 & $37\pm1$ & -7.3 \\
%0.62 & $16\pm1$ & -8.9 \\
%0.59 & $27\pm1$ & -6.0 \\
%0.56 & $14\pm2$ & -10.2

%%%%%%%%%%%%%%%%%%%%%%%%%%%%%%%%%%%%%%%%%%%%%%%%%%%%%%%%%%%%%%%%%%%%%%%%%

\end{document}